\documentclass[11pt]{article}
\usepackage{larecherche}

\def\be{\begin{equation}}
\def\ee{\end{equation}}
\def\bea{\begin{eqnarray}}
\def\eea{\end{eqnarray}}

\begin{document}
\vspace*{4cm} \title{Les composantes obscures de l'Univers: \\ La
  relativit\'e g\'en\'erale \`a l'\'epreuve des grandes \'echelles}

\author{Luc BLANCHET$^1$ \& Benoit FAMAEY$^2$} \address{$^1$ GRECO,
  Institut d'Astrophysique de Paris, CNRS UMR 7095, \\Universit\'e
  Pierre \& Marie Curie, 98$^{\rm bis}$ boulevard Arago, 75014 Paris,
  France\\ $^2$Observatoire astronomique de Strasbourg, CNRS UMR 7550,
  \\Universit\'e de Strasbourg, 11 rue de l'Universit\'e, 67000
  Strasbourg, France}

\maketitle

\abstracts{La cosmologie moderne, l'une des h\'eriti\`eres de notre
  th\'eorie de la gravitation, la relativit\'e g\'en\'erale, suppose
  l'existence dans l'Univers de grandes quantit\'es de mati\`ere noire
  et d'\'energie noire. Trouvera-t-on un jour la nature de ces deux
  composantes myst\'erieuses de l'Univers, ou bien refl\`etent-elles,
  au moins en partie, une limite \`a notre compr\'ehension de la
  gravitation\,?}

\section{L'av\`enement du mod\`ele standard de la cosmologie}

D\`es le XIX\`eme si\`ecle, l'application de la loi de la gravitation
universelle aux corps du syst\`eme solaire donna lieu \`a des
probl\`emes de ``masse manquante''. En effet, les orbites d'Uranus et
de Mercure \'etaient anormales, impliquant la pr\'esence \'eventuelle
d'autres plan\`etes non d\'etect\'ees jusqu'alors. Ces probl\`emes
furent r\'egl\'es de fa{\c c}on hybride, par une nouvelle plan\`ete
dans le cas de l'orbite d'Uranus -- Neptune, pr\'edite en 1846 par
Urbain Le Verrier et d\'ecouverte dans la foul\'ee -- et \textit{via}
une modification de la loi de la gravitation dans le cas de Mercure --
la relativit\'e g\'en\'erale d'Einstein en 1915.

A la fin de la premi\`ere moiti\'e du XX\`eme si\`ecle, des
probl\`emes relativement similaires de masse manquante commenc\`erent
\`a se poser aux \'echelles galactiques et extra-galactiques. A ces
\'echelles, le champ de gravitation est extr\^emement faible, et les
pr\'edictions dynamiques de la relativit\'e g\'en\'erale ne
diff\`erent gu\`ere de celles de la th\'eorie newtonienne. En 1933,
l'astronome suisse Fritz Zwicky~\cite{Zw33} mesura des vitesses
tr\`es \'elev\'ees pour quelques galaxies individuelles au sein de
l'amas de galaxies de Coma. En tenant compte de ces vitesses, il
calcula \`a partir des lois de Newton que la masse de l'amas devait
\^etre des centaines de fois plus importante que la masse
visible. Autrement dit, il devait exister dans l'amas de la mati\`ere
qu'on ne voit pas -- une masse invisible --, mais qui devrait
contribuer \`a sa dynamique. Il se trouve que la valeur qu'il avait
obtenue pour le rapport entre masse totale de l'amas et masse de
mati\`ere ordinaire \'etait largement surestim\'ee (notamment parce
qu'une large partie de la masse manquante \'etait sous la forme de gaz
chaud n'\'emettant qu'en rayons X), mais ce probl\`eme de masse
manquante n'a toujours pas enti\`erement disparu aujourd'hui.

Le probl\`eme soulev\'e par Zwicky ne fut n\'eanmoins pas pris au
s\'erieux pendant plusieurs d\'ecennies. La possibilit\'e de la
pr\'esence de grandes quantit\'es de mati\`ere invisible aux
\'echelles galactiques et au-del\`a revint en fait sur le devant de la
sc\`ene tout d'abord \`a cause de consid\'erations
\textit{th\'eoriques}. En effet, les premi\`eres simulations
num\'eriques de disques galactiques en rotation n'\'etant pas stables,
les chercheurs am\'ericains Jeremiah Ostriker et Jim
Peebles~\cite{OstPe} propos\`erent en 1973 que ces disques soient en
fait entour\'es de grands halos sph\'eriques de mati\`ere invisible
permettant de stabiliser les disques en rotation: de la mati\`ere
noire. M\^eme si on se rendit compte une dizaine d'ann\'ees plus tard
que cet argument th\'eorique n'\'etait pas tout \`a fait correct,
l'hypoth\`ese d'Ostriker et Peebles fut entre-temps corrobor\'ee par
la mesure de la vitesse de rotation du gaz dans les galaxies \`a
grandes distances de leur centre (gr\^ace \`a la raie \`a 21 cm de
transition hyperfine dans l'hydrog\`ene). Cette vitesse se r\'ev\'ela
approximativement constante en fonction de la distance au centre, une
d\'ecouverte observationnelle simultan\'ee des \'equipes des
astronomes am\'ericaine Vera Rubin et fran{\c c}ais Albert Bosma \`a
la fin des ann\'ees 70.~\cite{Rubin,Bosma} En effet, en l'absence de
mati\`ere noire, la vitesse de rotation aurait d\^u d\'ecro\^itre avec
le rayon. Aujourd'hui, on a besoin de cette mati\`ere noire pour
expliquer toute une s\'erie d'observations allant de l'\'echelle des
galaxies jusqu'au fond diffus cosmologique (voir
Annexe~\ref{annexe1}).

Par ailleurs, la pr\'esence de ces quantit\'es importantes de
mati\`ere noire aux grandes \'echelles fut rapidement consid\'er\'ee
comme permettant \`a l'Univers d'atteindre la densit\'e critique
d'\'energie pour laquelle l'Univers est spatialement plat \`a grande
\'echelle. La platitude de l'Univers \'etait en effet naturelle dans
le cadre de la th\'eorie de l'inflation initiale de l'expansion
(br\`eve phase d'expansion acc\'el\'er\'ee au sortir de l'\`ere de
Planck), propos\'ee en particulier par Alan Guth au d\'ebut des
ann\'ees 80 pour expliquer notamment l'homog\'en\'eit\'e initiale de
l'Univers. Ceci mena \`a un premier mod\`ele ``standard'' de la
cosmologie dans les ann\'ees 80--90, pour lequel la densit\'e totale
de mati\`ere dans l'Univers \'etait la densit\'e critique, mais la
fraction de mati\`ere ordinaire de l'ordre de seulement quelques
pourcents.

Cependant, ce mod\`ele fut rapidement confront\'e \`a des
probl\`emes. Par exemple, la fraction de mati\`ere ordinaire
observ\'ee dans les amas de galaxies \'etait significativement plus
importante que les quelques pourcents attendus. Ceci amena Jeremiah
Ostriker et Paul Steinhardt~\cite{OS95} \`a proposer en 1995 un
nouveau mod\`ele de la cosmologie, domin\'e par une \textit{constante
  cosmologique}, identique \`a la constante propos\'ee par Einstein en
1917 pour g\'en\'eraliser ses \'equations de la relativit\'e
g\'en\'erale. Ainsi la fraction de mati\`ere ordinaire pouvait rester
de seulement quelques pourcents en terme d'\'energie totale, domin\'ee
par la constante cosmologique, mais \'etait bien plus importante
qu'avant en terme de fraction de la mati\`ere. 

Une pr\'ediction importante d'un tel mod\`ele avec constante
cosmologique positive \'etait une acc\'el\'eration de l'expansion de
l'Univers, qui fut confirm\'ee d\`es 1998 avec l'observation de
supernov{\ae} lointaines par les \'equipes de Brian Schmidt, Adam
Riess et Saul Perlmutter, qui re{\c c}urent le prix Nobel en 2011 pour
ces observations. La constante cosmologique, dont l'origine reste
myst\'erieuse aujourd'hui, peut \^etre assimil\'ee \`a un fluide
exotique avec une pression n\'egative, donc qui ``antigravite'', et
qui ne se dilue pas dans l'expansion de l'espace-temps. Ce fluide
myst\'erieux est souvent appel\'e l'\'energie noire, suppos\'ee
n'avoir aucun lien avec la mati\`ere noire.

\section{Les succ\`es du mod\`ele standard de la cosmologie}

Telle est la situation en ce d\'ebut de XXI\`eme si\`ecle: le mod\`ele
standard de la cosmologie d\'ecrit l'expansion de l'Univers \`a partir
d'une phase d'inflation initiale suivie d'une phase d'expansion plus
mod\'er\'ee, mais qui est en train de recommencer \`a acc\'el\'erer
sous l'effet de la constante cosmologique. Le contenu de l'Univers se
r\'epartit entre \'energie noire (la constante cosmologique),
mati\`ere noire, et mati\`ere ordinaire (les ``baryons'' dont
nous-m\^emes et toutes les \'etoiles sommes constitu\'es). Les
param\`etres de ce mod\`ele ont \'et\'e progressivement pr\'ecis\'es
gr\^ace \`a l'observation des grandes structures, et surtout par
l'analyse r\'ecente des satellites WMAP et Planck du fond diffus
cosmologique, la premi\`ere lumi\`ere que nous recevons de
l'Univers, \'emise 380\,000 ans apr\`es le Big Bang (voir
Annexe~\ref{annexe1}).

On peut dire que les fluctuations de la temp\'erature du fond diffus
cosmologique et leur ajustement par le mod\`ele standard ont
transform\'e la cosmologie en une science de pr\'ecision. La hauteur
des pics de fluctuations de temp\'erature nous informe sans
\'equivoque de la pr\'esence de mati\`ere noire aux \'echelles
cosmologiques, se comportant comme un fluide de particules sans
dissipation. De plus cette mati\`ere noire a fa{\c c}onn\'e l'Univers
tel que nous le connaissons actuellement. La formation des structures
(amas de galaxies, grandes structures, \textit{etc.}) dans ce mod\`ele
est r\'ealis\'ee gr\^ace \`a l'attraction gravitationnelle engendr\'ee
par la mati\`ere noire. C'est l'un des grands succ\`es du mod\`ele
standard que d'expliquer la formation et l'\'evolution des
grandes structures observ\'ees dans l'Univers. Celles-ci forment une
toile d'araign\'ee avec des filaments gigantesques compos\'es de
milliers de galaxies, joints entre eux par des super-amas de galaxies
et de gaz chaud.  Ces grandes structures sont maintenant reproduites
avec pr\'ecision par des simulations num\'eriques dans le cadre du
mod\`ele standard.

Le pourcentage mesur\'e de mati\`ere ordinaire par rapport \`a la
mati\`ere noire est en accord avec la synth\`ese des
\'el\'ements l\'egers qui eut lieu quelques centaines de secondes
apr\`es le Big Bang. A cet instant la temp\'erature de l'Univers
descend au-dessous du milliard de degr\'es et les protons et neutrons
peuvent alors se combiner pour former des noyaux atomiques l\'egers
comme le deut\'erium et l'h\'elium. La synth\`ese de l'h\'elium
  dans l'Univers primitif et la pr\'ediction du taux correct de
production avaient constitu\'e l'une des grandes confirmations du Big
Bang avec la d\'ecouverte du rayonnement cosmologique en 1965. Le
mod\`ele standard cosmologique a confirm\'e et \'etendu ces succ\`es
initiaux en prouvant la coh\'erence de la mesure du rapport entre
mati\`ere noire et mati\`ere ordinaire avec les fluctuations du fond
cosmologique.

Les d\'ecomptes du nombre de galaxies ont permis de mettre en
\'evidence une autre pr\'ediction du mod\`ele standard: des ondes dans
la distribution spatiale des galaxies, avec une longueur d'onde de
l'ordre de 100 Mpc, qui ne sont autres que l'empreinte des
oscillations acoustiques dans le plasma \'electron-proton primordial,
et qui ont conduit aux fluctuations observ\'ees du fond diffus
cosmologique. Ces ondes acoustiques baryoniques (appel\'ees BAO)
furent d\'etect\'ees pour la premi\`ere fois en 1999.

Le mod\`ele cosmologique poss\`ede une simplicit\'e et une coh\'erence
qui en font la base pour \'etudier de nombreux ph\'enom\`enes
importants, et il permet d'avoir un acc\`es \`a la cosmologie
primordiale comme la p\'eriode de l'inflation. Il repose
  n\'eanmoins sur un certain nombre d'hypoth\`eses fortes:
\begin{itemize}
\item La relativit\'e g\'en\'erale est valable \`a toutes les
  \'echelles (jusqu'\`a l'\'echelle de Planck);
\item Nous n'occupons pas une place privil\'egi\'ee dans l'Univers
  (principe dit ``copernicien'');
\item Les effets de mati\`ere noire ``cosmologique'', n\'ecessaire
  pour reproduire les fluctuations du fond diffus, et les effets \`a
  l'\'echelle des galaxies, pour reproduire les courbes de rotation,
  proviennent des m\^emes causes;
\item L'\'energie noire est une entit\'e distincte de la mati\`ere
  noire.
\end{itemize}
La validit\'e de certaines de ces hypoth\`eses pourrait, comme nous
allons le voir, \^etre remise en question.

\section{Probl\`emes th\'eoriques et observationnels dans le mod\`ele standard}

L'existence du secteur sombre (mati\`ere noire et \'energie noire)
r\'ev\`ele un probl\`eme de compatibilit\'e entre la physique
fondamentale connue et les observations cosmologiques. Aucune
particule connue dans le cadre du mod\`ele standard actuel de la
physique des particules ne peut jouer le r\^ole de la particule de
mati\`ere noire (voir Annexe~\ref{annexe2}). L'extension de ce
mod\`ele standard de physique des particules pour rendre compte de
l'existence de mati\`ere noire ne constitue n\'eanmoins pas un
probl\`eme th\'eorique \textit{a priori}. Cependant, le vrai
probl\`eme est qu'on cherche ces particules de mati\`ere noire depuis
des ann\'ees, et qu'on n'en a toujours pas d\'etect\'ees, malgr\'e de
nombreuses fausses alertes. Ni directement par des exp\'eriences en
laboratoire -- soit des capteurs de ces particules sur Terre soit des
exp\'eriences de recherche sur les acc\'el\'erateurs de particules --
ni indirectement en Astronomie par des observations qui seraient
expliqu\'ees par l'annihilation possible de la mati\`ere noire en
d'autres particules. L'espace de param\`etres des extensions
naturelles du mod\`ele standard de physique des particules se r\'eduit
donc de jour en jour, et l'absence persistante de d\'etection de la
mati\`ere noire engendrerait sur le long terme une crise profonde du
domaine.

Le deuxi\`eme probl\`eme du mod\`ele standard de la cosmologie est
li\'e \`a la constante cosmologique. Celle-ci est ajout\'ee \`a la
main dans les \'equations d'Einstein comme param\`etre
suppl\'ementaire, et appara\^it comme naturelle d'un point de vue
relativiste. En fait, comme rien ne l'interdit, et qu'en physique on
aime le ``principe totalitaire'' selon lequel tout ce qui n'est pas
interdit est obligatoire, la constante cosmologique doit bien, en
principe, \^etre incluse dans notre description du champ
gravitationnel. Il suffit de v\'erifier qu'elle reste compatible avec
les tests de la relativit\'e g\'en\'erale dans le syst\`eme solaire et
les pulsars binaires. Mais l\`a n'est pas le probl\`eme. En effet la
valeur \textit{mesur\'ee} de la constante cosmologique est
extr\^emement petite, elle vaut environ $10^{-123}$ en unit\'es de
Planck. Le probl\`eme intervient lorsque l'on essaie de comprendre
cette valeur en utilisant la th\'eorie quantique des champs. Comme l'a
montr\'e Andre\"i Sakharov en 1968 (\textit{le} Sakharov des droits de
l'homme), l'\'energie des fluctuations quantiques du vide en th\'eorie
des champs prend n\'ecessairement la forme d'une constante
cosmologique. C'est pourquoi cette constante est souvent assimil\'ee
\`a l'\'energie du vide. En th\'eorie des champs, quand on cherche \`a
d\'ecrire des syst\`emes microscopiques, on ignore g\'en\'eralement
l'\'energie du vide car on ne s'int\'eresse qu'\`a des diff\'erences
d'\'energie et que celle-ci dispara\^it dans la diff\'erence. Mais en
relativit\'e g\'en\'erale, d'apr\`es le principe d'\'equivalence,
toutes les formes d'\'energies sont la source du champ
gravitationnel. L'\'energie du vide n'est qu'une forme d'\'energie
particuli\`ere et il faut donc la mettre dans les \'equations
d'Einstein au m\^eme titre que les autres formes d'\'energie. On doit
alors imaginer que tous les champs pr\'esents dans l'Univers
contribuent \`a cette \'energie du vide. Lorsque l'on fait le calcul
de l'\'energie du vide pour un champ donn\'e, on trouve une valeur
gigantesque, qui n'a rien \`a voir avec la minuscule valeur
observ\'ee. Et en plus il faut additionner toutes les contributions de
tous les champs connus, ce qui donne une valeur encore plus
titanesque. Ce r\'esultat manifestement absurde a \'et\'e appel\'e la
pr\'ediction la plus fausse de toute l'histoire de la physique
th\'eorique\,! La n\'ecessit\'e de devoir compenser it\'erativement
l'\'energie du vide pour la rendre finalement minuscule sans l'annuler
totalement est ce qu'on appelle un probl\`eme ``d'ajustement fin''
(\textit{fine-tuning}). Par ailleurs, le fait que les densit\'es
d'\'energie noire, de mati\`ere noire et de mati\`ere ordinaire
soient, \`a peu de choses pr\`es, du m\^eme ordre de grandeur
aujourd'hui, n'est pas non plus compris: c'est un probl\`eme dit ``de
co\"incidence'', qui pourrait pointer vers une physique plus complexe
que suppos\'ee dans le mod\`ele standard.

\begin{figure}[htb]
\centerline{\includegraphics[width=10cm]{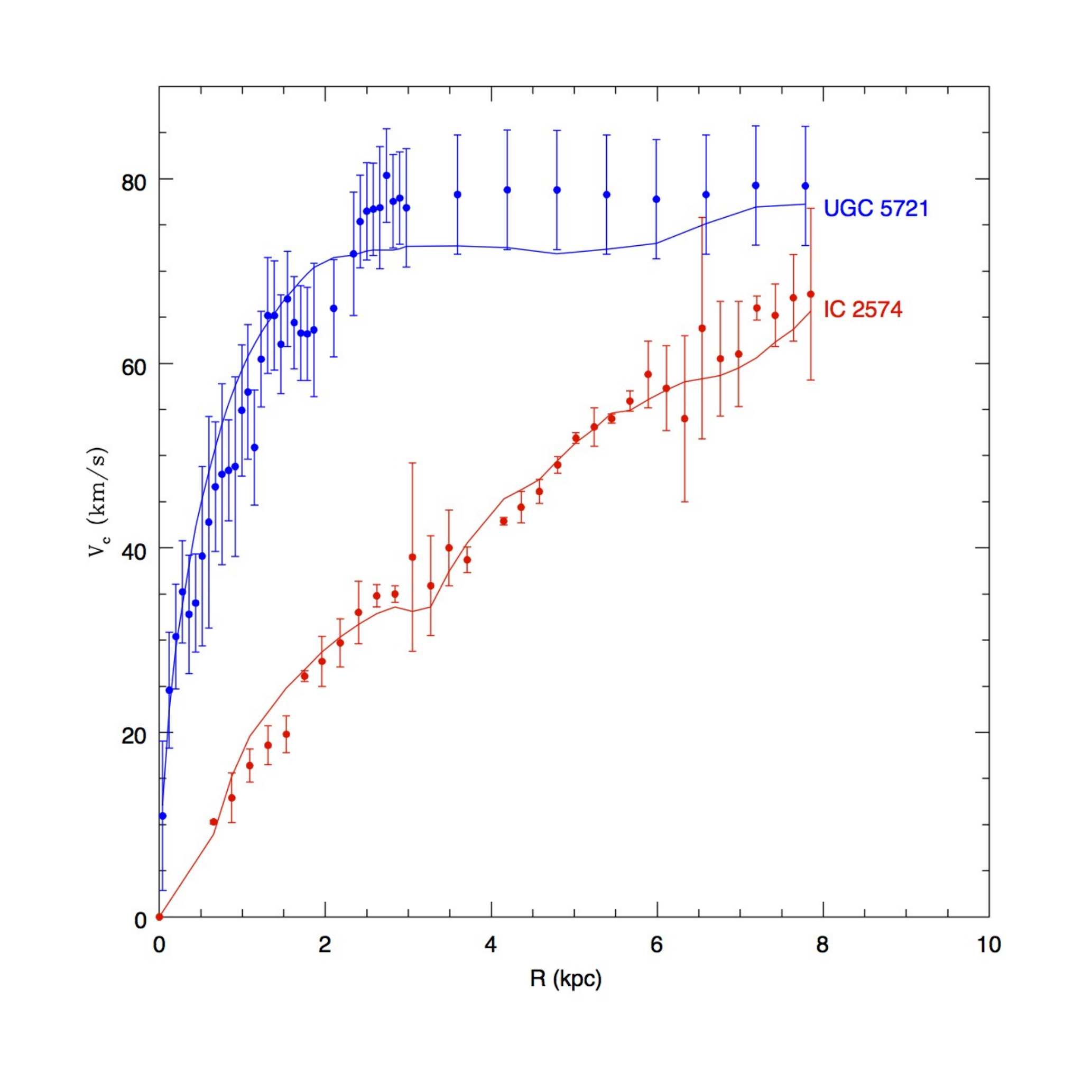}}
\vspace{-1cm} 
\caption{Courbes de rotation, c'est-\`a-dire vitesse circulaire du gaz
  (en km/s) en fonction du rayon (en kpc), dans les galaxies UGC~5721
  et IC~2574 (les points indiquent les mesures de la vitesse
  accompagn\'ees de leurs barres d'erreur): ces deux galaxies ont
  quasiment la m\^eme masse totale de mati\`ere noire, mais leur
  \textit{distribution} de mati\`ere ordinaire (gaz et \'etoiles)
  diff\`ere. Dans le mod\`ele standard, on s'attendrait \textit{a
    priori} \`a ce que les formes des courbes de rotation soient \`a
  peu de choses pr\`es les m\^emes. Or ce n'est pas du tout le cas. La
  pr\'ediction de MOND pour ces courbes de rotation est quant \`a elle
  repr\'esent\'ee en lignes continues: la forme des courbes de
  rotation est parfaitement pr\'edite dans le paradigme MOND.}
\label{fig1}
\end{figure}

En plus des probl\`emes th\'eoriques pr\'ec\'edents, le mod\`ele
standard est aussi confront\'e \`a des probl\`emes observationnels,
principalement \`a l'\'echelle des galaxies. En supposant que l'on
peut extrapoler le mod\`ele depuis l'\'echelle cosmologique, o\`u il
marche fort bien, jusqu'aux \'echelles galactiques gr\^ace aux
simulations num\'eriques, de nombreux probl\`emes apparaissent. Par
exemple, on ne comprend absolument pas les propri\'et\'es internes et
la distribution des galaxies satellites orbitant autour des grandes
galaxies spirales du Groupe Local, notre Voie Lact\'ee et la galaxie
d'Androm\`ede.~\cite{Pawlowski} De plus, dans toutes les galaxies
spirales, on observe une forte corr\'elation entre la distribution de
mati\`ere ordinaire et le champ gravitationnel d\'eduit de la courbe
de rotation des galaxies, alors que l'on s'attendrait plut\^ot
\textit{a priori} \`a une corr\'elation entre la distribution de
mati\`ere noire et le champ gravitationnel, puisque la mati\`ere noire
est cens\'ee dominer la masse des galaxies. Plus pr\'ecis\'ement, si
on ignore la mati\`ere ordinaire, n\'egligeable dans un certain nombre
de petites galaxies, un halo de mati\`ere noire de masse donn\'ee
devrait toujours engendrer le m\^eme profil de courbe de rotation. Or
on observe une grande diversit\'e de profils de courbes de
rotation~\cite{Oman} \`a masse totale donn\'ee (voir
Figure~\ref{fig1}). Dans le cadre standard, cette diversit\'e peut
\textit{a priori} s'expliquer si les effets produits par exemple par
les explosions de supernov{\ae} d\'eplacent de grandes quantit\'es de
mati\`ere ordinaire dont l'effet gravitationnel modifie alors le
profil de mati\`ere noire. C'est ce qu'on appelle un m\'ecanisme de
r\'etroaction (\textit{feedback}). Des effets de r\'etroaction
diff\'erents selon les galaxies peuvent alors expliquer la diversit\'e
des profils \`a masse totale donn\'ee. Mais comment alors expliquer
l'\textit{uniformit\'e}~\cite{FamMcG12} de ces profils en fonction de
la distribution de mati\`ere ordinaire\,?  Cela n\'ecessite un
ajustement fin des param\`etres de r\'etroaction (par exemple, les
m\'ecanismes de r\'etroaction internes doivent finement balancer les
effets li\'es \`a l'environnement et \`a l'histoire des
  collisions et fusions mineures et majeures de chaque galaxie) --
ajustement fin qui n'est pas du tout compris aujourd'hui dans le cadre
standard.

\textit{A contrario}, si on suppose que l'effet que l'on attribue \`a
la mati\`ere noire est en fait d\^u \`a une modification effective de
la gravitation \`a l'\'echelle des galaxies, qui impliquerait
l'existence d'une nouvelle constante sous la forme d'une
acc\'el\'eration caract\'eristique, alors tout s'explique (voir
Figure~\ref{fig1}). Bien s\^ur ceci reste tr\`es sp\'eculatif en
l'absence d'une fa{\c c}on pr\'ecise pour mettre en {\oe}uvre une
telle constante dans les lois de la nature, particuli\`erement la loi
gravitationnelle. Mais il se trouve qu'il existe une formule simple
propos\'ee par le physicien isra\'elien Mordehai Milgrom en
1983~\cite{Milg1,Milg2,Milg3} (connue sous le nom de MOND -- MOdified
Newtonian Dynamics) qui modifie la loi de la gravitation newtonienne
pour des champs tr\`es faibles, en de{\c c}\`a d'une acc\'el\'eration
caract\'eristique (voir Annexe~\ref{annexe3}). Autrement dit, avec
cette loi, pour les acc\'el\'erations tr\`es faibles, la gravitation
peut devenir beaucoup plus intense que celle de Newton. Cette formule
est extr\^emement pr\'edictive et marche remarquablement bien pour
reproduire toutes les observations connues sur la mati\`ere noire au
niveau des galaxies.~\cite{FamMcG12} Myst\'erieusement, la valeur
mesur\'ee de la constante d'acc\'el\'eration est tr\`es proche de la
valeur associ\'ee \`a la constante cosmologique, montrant qu'il existe
peut-\^etre un lien entre l'\'energie noire et les effets que l'on
attribue g\'en\'eralement \`a la mati\`ere noire.

Malheureusement, la formule MOND a ses propres probl\`emes car elle ne
fonctionne pas partout, loin de l\`a: malgr\'e son incroyable
efficacit\'e dans les galaxies, elle \'echoue \`a plus grande
\'echelle pour d\'ecrire les amas de galaxies.~\cite{GD92,Clowe06}
Elle est aussi mise \`a l'\'epreuve \`a petite \'echelle pour
expliquer les mouvements des \'etoiles dans certains petits amas
stellaires. En outre, elle n'est pas utilisable en
cosmologie. Enfin, \`a l'\'echelle du syst\`eme solaire,
  certaines transitions entre les r\'egimes de forte et faible
  acc\'el\'eration ne sont pas permises par les mouvements
  plan\'etaires, m\^eme si d'autres restent parfaitement acceptables~\cite{Hees}.

\section{Alternatives au mod\`ele standard et th\'eories de gravitation modifi\'ee}

Face \`a ces probl\`emes th\'eoriques et observationnels du mod\`ele
standard de la cosmologie, de nombreuses alternatives sont
consid\'er\'ees par la communaut\'e. Comme le mod\`ele repose sur
l'hypoth\`ese de la validit\'e de la relativit\'e g\'en\'erale, il est
naturel de consid\'erer des th\'eories de la gravitation
modifi\'ees. Ces derni\`eres ann\'ees on a assist\'e \`a une floraison
impressionnante de telles th\'eories.

Il y a tout d'abord les th\'eories qui cherchent \`a interpr\'eter
l'\'energie noire -- une simple constante cosmologique dans le
mod\`ele standard -- par un champ fondamental nouveau. Ce champ est
g\'en\'eralement baptis\'e ``quintessence'', car il peut \^etre vu
comme une cinqui\`eme interaction fondamentale se rajoutant aux quatre
forces connues (gravitation, \'electromagn\'etisme, interactions forte
et faible). Dans ce cas, la gravitation modifi\'ee consiste \`a
rajouter au champ tensoriel habituel de la relativit\'e g\'en\'erale
-- le champ qui repr\'esente la m\'etrique, l'objet math\'ematique qui
fixe les distances dans l'espace-temps -- un champ scalaire de
quintessence. C'est pourquoi on nomme cette extension de la
relativit\'e g\'en\'erale une th\'eorie ``tenseur-scalaire''.

Les th\'eories tenseur-scalaire sont testables exp\'erimentalement,
car dans ce cadre, l'\'energie noire n'est plus une simple constante
cosmologique, mais un champ qui varie avec l'espace et le
temps. Ainsi, \`a partir de 2020, le satellite europ\'een Euclid
pourra v\'erifier, en \'evaluant pr\'ecis\'ement le taux d'expansion
de l'Univers en fonction du temps, si celui-ci est compatible avec une
constante cosmologique ou avec un champ de quintessence. Une autre
classe de mod\`eles tr\`es en vogue actuellement sont les th\'eories
dites ``$f(R)$'', qui remplacent le terme de scalaire de courbure
``$R$'' de l'action d'Einstein-Hilbert de la relativit\'e g\'en\'erale
par une fonction de celui-ci suppos\'ee \^etre contrainte par les
observations cosmologiques.

Un probl\`eme important avec les champs scalaires utilis\'es pour
d\'ecrire l'\'energie noire est de satisfaire les contraintes
observationnelles dans le syst\`eme solaire et les pulsars binaires,
donc dans les r\'egions o\`u il y a sur-densit\'e de la mati\`ere
ordinaire par rapport \`a la densit\'e moyenne cosmologique. En effet,
l'analyse pr\'ecise de la trajectoire des sondes spatiales et des
plan\`etes dans le syst\`eme solaire, ainsi que le chronom\'etrage des
pulsars binaires restreint les th\'eories de ce type. Une th\'eorie
tenseur-scalaire particuli\`ere a \'et\'e invent\'ee pour s'affranchir
des contraintes dans le syst\`eme solaire. Le champ scalaire s'appelle
le ``cam\'el\'eon'' car il s'adapte et se camoufle comme un
cam\'el\'eon dans certains environnements comme le syst\`eme solaire
o\`u il devient ind\'etectable.~\cite{KW04} L'imagination cr\'eatrice
des th\'eoriciens est sans limite\,!

Un autre type de th\'eorie de gravitation modifi\'ee est sous les feux
de la rampe depuis quelques ann\'ees. Dans sa formulation
g\'eom\'etrique, la relativit\'e g\'en\'erale est interpr\'et\'ee
comme r\'esultant d'une d\'eformation de l'espace-temps, mais on peut
aussi voir la force gravitationnelle comme \'etant associ\'ee \`a
l'\'echange d'une particule, le graviton. En relativit\'e
g\'en\'erale, le graviton a une masse rigoureusement nulle (comme le
photon qui a aussi une masse nulle, mais un spin \'egal \`a 1 alors
que le graviton a un spin 2). Dans les ann\'ees 1930, les physiciens
suisse Markus Fierz et autrichien Wolfgang Pauli ont cherch\'e \`a
savoir s'il serait possible que, dans le cadre d'une extension de la
relativit\'e g\'en\'erale, la force gravitationnelle soit
v\'ehicul\'ee par un graviton massif.

La r\'eponse \`a cette question a subi des vicissitudes jusqu'au
d\'ebut des ann\'ees 2010 quand l'\'equipe de la physicienne suisse
Claudia de Rham, de l'universit\'e Case Western Reserve \`a Cleveland,
a propos\'e une extension de la gravitation avec un graviton
massif.~\cite{dRGT10} Contrairement aux th\'eories pr\'ec\'edentes du
m\^eme type, qui contenaient des \'etats d'excitation instables
appel\'es ``fant\^omes'' qui rendent la th\'eorie inexploitable (et
ont tendance \`a prolif\'erer dans les th\'eories alternatives), il
s'agit d'une ``bonne'' th\'eorie. De plus, cette th\'eorie peut \^etre
\'etendue \`a une th\'eorie dite ``bim\'etrique'', dans laquelle il y
a un graviton massif et un graviton sans masse, et qui repr\'esente
l'unique extension de la relativit\'e g\'en\'erale avec deux
m\'etriques et, bien s\^ur, sans fant\^omes. Pour tenter de se
repr\'esenter une telle th\'eorie, on peut imaginer qu'aux deux
gravitons sont associ\'es deux espace-temps superpos\'es d\'ecrits
chacun par une m\'etrique diff\'erente. On pourrait avoir de la
mati\`ere ordinaire dans l'un des deux espace-temps et une mati\`ere
``exotique'' dans l'autre espace-temps, ce qui se traduirait par de
nouvelles propri\'et\'es pour le mouvement de la mati\`ere exotique
tel que mesur\'e dans l'espace-temps ordinaire.

En quoi cette th\'eorie de la gravitation massive pourrait-elle
r\'esoudre les probl\`emes cosmologiques\,? Dans cette th\'eorie, la
masse du graviton pourrait \^etre associ\'ee \`a une constante
cosmologique de sorte qu'on esp\`ere que cette masse donnerait la
valeur de l'\'energie noire. Il devrait ainsi \^etre plus facile de
r\'esoudre le probl\`eme de la valeur incroyablement petite mais non
nulle de la constante cosmologique. Comme la gravitation massive
est une th\'eorie r\'ecente, ses pr\'edictions n'ont pas encore
\'et\'e totalement \'eclaircies. On ignore si la
cosmologie est totalement satisfaisante dans le cadre de cette
th\'eorie.

D'autres th\'eories de la gravitation modifi\'ee sont motiv\'ees quant
\`a elles par le probl\`eme de la mati\`ere noire. Elles ont
l'ambition de retrouver la formule MOND qui d\'ecrit si bien le
champ gravitationnel attribu\'e \`a la mati\`ere noire dans les galaxies. En effet,
rappelons que cette formule ne constitue pas \`a proprement parler une th\'eorie (en
particulier elle n'est pas relativiste), mais repr\'esente
une loi empirique dont on constate qu'elle marche extr\^emement bien dans un
  certain r\'egime, mais qu'on ne comprend pas en termes de physique
fondamentale.

L'id\'ee des th\'eories inspir\'ees par MOND consiste en g\'en\'eral
\`a rajouter des champs nouveaux au champ m\'etrique de la
relativit\'e g\'en\'erale (champs scalaires et vectoriels) de fa{\c
  c}on \`a n'avoir plus besoin de mati\`ere noire. En particulier, la
th\'eorie relativiste dite TeVeS pour ``tenseur-vecteur-scalaire'',
d\'evelopp\'ee par les physiciens isra\'elien Jacob Bekenstein et
am\'ericain Bob Sanders,~\cite{Bek04,Sand05} a jou\'e un r\^ole de
pionnier dans ce domaine. De nombreuses autres th\'eories ont \'et\'e
propos\'ees depuis lors, mais la plupart sont compliqu\'ees et toutes
contiennent une fonction arbitraire mise \`a la main dans l'``action''
de la th\'eorie -- ce qui n'est pas tr\`es satisfaisant. De plus
certaines de ces th\'eories sont de vraies maisons hant\'ees car
remplies de fant\^omes\,! Enfin toutes ces th\'eories, qui sont sans
mati\`ere noire, ont beaucoup de difficult\'es \`a reproduire les
observations des fluctuations du fond diffus cosmologique
mesur\'ees par le satellite Planck.

Peut-\^etre que ces th\'eories p\^atissent d'un fort pr\'ejudice
th\'eorique, qui est en fait ancr\'e dans l'esprit original de MOND:
trouver une th\'eorie parfaite de gravitation modifi\'ee sans
mati\`ere noire, valable \`a toutes les \'echelles. Mais apr\`es tout,
les observations montrent clairement la pr\'esence d'un nouveau
degr\'e de libert\'e se comportant comme un fluide sans dissipation
aux grandes \'echelles, donc comme de la mati\`ere noire. D'un autre
c\^ot\'e, il y a \textit{dans les donn\'ees} aux \'echelles
galactiques quelque chose qui ressemble \`a une loi nouvelle --
peut-\^etre un m\'ecanisme de gravit\'e modifi\'ee.

Une th\'eorie hybride (sans pr\'ejudice), avec de la mati\`ere noire
se couplant \`a une physique gravitationnelle modifi\'ee,
pourrait-elle alors rendre compte de toutes les observations
connues\,? Dans cet esprit une forme nouvelle de mati\`ere noire a
\'et\'e propos\'ee. Cette mati\`ere noire est fond\'ee sur une
analogie avec la physique des mat\'eriaux di\'electriques en
\'electrostatique. Dans un mat\'eriau di\'electrique, les atomes se
comportent comme des dip\^oles \'electriques, qui se polarisent en
pr\'esence d'un champ \'electrique ext\'erieur: ils s'orientent
suivant le champ et deviennent eux-m\^emes la source d'un nouveau
champ \'electrique. Le champ total est alors la somme du champ
ext\'erieur et du champ induit par la polarisation des
atomes. Remarquablement, il se trouve que MOND appara\^it comme une
sorte d'analogue gravitationnel au ph\'enom\`ene de polarisation. Dans
le cas gravitationnel, la polarisation entra\^ine l'augmentation de la
force de gravitation, d'o\`u la la proposition d'un nouveau type de mati\`ere
  noire qui est d\'ecrite par sa masse et est munie d'un moment
dipolaire ``gravitationnel'' -- l'analogue gravitationnel du moment
dipolaire \'electrique. La ph\'enom\'enologie de cette mati\`ere noire
est alors exactement la m\^eme que celle de MOND.

Tr\`es r\'ecemment, cette id\'ee a conduit \`a plusieurs th\'eories
relativistes de mati\`ere noire ``dipolaire''. L'avantage dans ce type
de th\'eories, est que comme il y a de la mati\`ere noire on peut
esp\'erer \^etre en accord avec les observations cosmologiques. Dans
la derni\`ere version de ces th\'eories, propos\'ee par l'un de nous
(L.B.) avec Lavinia Heisenberg, de l'universit\'e de
Zurich,~\cite{BH15b} la gravitation est aussi modifi\'ee et se place
dans le cadre bien pos\'e (sans fant\^omes) des th\'eories
``bim\'etriques'' de la gravit\'e massive mentionn\'ees
pr\'ec\'edemment. Il s'agit donc d'une th\'eorie hybride qui suppose
la pr\'esence d'une mati\`ere noire particuli\`ere (de nature
dipolaire) coupl\'ee \`a une extension de la relativit\'e
g\'en\'erale. Une telle th\'eorie pourrait r\'esoudre \`a la fois le
probl\`eme de l'\'energie noire par la gravit\'e modifi\'ee, et celui
de la mati\`ere noire par des particules de type dipolaire se
comportant comme un fluide sans dissipation \`a l'\'echelle du fond
diffus cosmologique, et reproduisant MOND aux \'echelles
galactiques.

Toutes ces th\'eories refl\`etent une recherche tout \`a fait
pertinente dans l'\'etat actuel de nos connaissances, et \'etant
donn\'es les r\'esultats exp\'erimentaux et observationnels dont nous
disposons \`a ce jour. Mais si par exemple on d\'ecouvrait des
particules de mati\`ere noire ``normale'' (WIMP, neutrinos st\'eriles,
\textit{etc.}) par des exp\'eriences sur Terre (dans des capteurs ou
avec les acc\'el\'erateurs de particules comme le LHC), il est
probable que tout ce champ de recherche de m\^eme que l'id\'ee
g\'en\'erale de MOND seraient s\'erieusement \`a remettre en cause. Il
resterait n\'eanmoins \`a comprendre pourquoi mati\`ere noire et
mati\`ere ordinaire conspirent \`a l'\'echelle des galaxies de fa{\c
  c}on \`a reproduire la ph\'enom\'enologie de MOND, qui est un simple
fait observationnel qu'il nous faut comprendre, quel que soit le cadre
th\'eorique.

\section{Conclusion}

Nous disposons aujourd'hui d'un excellent mod\`ele standard de la
cosmologie, qui d\'ecrit l'Univers aux grandes \'echelles avec une
tr\`es grande pr\'ecision, mais qui n'explique pas l'origine et la
nature des constituants fondamentaux de l'Univers (mati\`ere noire et
\'energie noire), et qui n'est pas encore capable d'expliquer les
observations de la dynamique des galaxies. Ce dernier point est
souvent consid\'er\'e comme potentiellement soluble par une meilleure
prise en compte de la physique complexe de la mati\`ere ordinaire dans
les simulations de formation des galaxies. Il est n\'eanmoins probable
que ces probl\`emes vont persister, car ils font peut-\^etre poindre
une physique fondamentale plus complexe de la mati\`ere noire,
coupl\'ee \`a d'\'eventuelles modifications de la relativit\'e
g\'en\'erale, et qui pourrait \'egalement illuminer d'un jour nouveau
le probl\`eme de l'\'energie noire. Il reste qu'aujourd'hui les
mod\`eles alternatifs que nous avons \'evoqu\'es ne peuvent pas
v\'eritablement rivaliser avec le mod\`ele standard. Ils doivent avant
tout \^etre valid\'es par une physique plus fondamentale avant de
pouvoir faire des pr\'edictions nouvelles, \'eventuellement hors du
contexte cosmologique. Les ann\'ees \`a venir seront, quoi qu'il en
soit, passionnantes.

\appendix

\section{Le fond diffus cosmologique}
\label{annexe1}

Le fond diffus cosmologique est un rayonnement fossile, vestige de
l'\'epoque dense et chaude qu'a connue l'Univers dans le pass\'e juste
apr\`es le Big Bang, d\'ecouvert fortuitement par Arno Penzias et
Robert Wilson en 1965. Lorsque les protons et les \'electrons se sont
recombin\'es pour former les premiers atomes, 380\,000 ans apr\`es le
Big Bang, les photons ont \'et\'e lib\'er\'es et se sont propag\'es
jusqu'\`a nous. Les fluctuations du fond diffus cosmologique, qui
correspondent aux variations de densit\'e initiales de l'Univers \`a
cette \'epoque ont \'et\'e d\'etect\'ees en 1992, et constituent
une mine d'informations extraordinaire sur l'\'etat de l'Univers au
moment o\`u le rayonnement a \'et\'e \'emis. Ces fluctuations sont les
``graines'' qui engendrent les grandes structures de l'Univers et plus
tard les galaxies. Une d\'ecouverte plus r\'ecente a montr\'e que l'on
peut retrouver ces graines sous la forme d'ondes de densit\'e dans la
distribution spatiale des galaxies. La mati\`ere noire est
indispensable pour expliquer tant l'amplitude des fluctuations
elles-m\^emes que la formation et l'\'evolution des grandes
structures, car c'est elle qui d\'eclenche et amplifie l'effondrement
gravitationnel de la mati\`ere ordinaire. Le spectre complet des
fluctuations du fond diffus cosmologique est parfaitement expliqu\'e
par le mod\`ele cosmologique standard, comme les r\'esultats r\'ecents
du satellite Planck l'ont confirm\'e. L'ajustement de ce mod\`ele aux
observations implique un Univers compos\'e aujourd'hui de 68\%
d'\'energie noire, de 27\% de mati\`ere noire et -- seulement -- de
5\% de mati\`ere ordinaire. M\^eme si \'energie noire et mati\`ere
noire sont \`a ce stade des mots pour ce que nous ne comprenons pas
encore\,!

\section{A la recherche des particules de mati\`ere noire}
\label{annexe2}

Les particules connues actuellement en physique des particules ne
  peuvent pas \^etre la particule de mati\`ere noire. Mais lorsque
l'on examine des extensions du mod\`ele actuel de la physique
des particules, les candidats \`a la mati\`ere noire ne manquent
pas\,! Le meilleur est le WIMP (Weakly Interacting Massive Particle),
une particule massive qui n'interagit avec la mati\`ere ordinaire que
\textit{via} l'interaction faible. La particule stable la plus
l\'eg\`ere dans le cadre d'une extension supersym\'etrique du mod\`ele
des particules est un WIMP appel\'e neutralino. D'autres candidats
int\'eressants sont l'axion, une particule invoqu\'ee pour r\'esoudre
certains probl\`emes de physique des particules (n'ayant \textit{a
  priori} rien \`a voir avec le probl\`eme de la mati\`ere noire), les
neutrinos st\'eriles qui seraient similaires aux neutrinos du mod\`ele
standard mais sans interactions faibles et avec des masses plus
\'elev\'ees, et les particules de Kaluza-Klein qui seraient des
\'etats excit\'es des particules du mod\`ele standard dans des
th\'eories ayant des dimensions d'espace suppl\'ementaires.  Toutes
ces particules candidates pour la mati\`ere noire sont recherch\'ees
activement, mais aucune n'a \'et\'e d\'etect\'ee \`a ce jour.

\section{MOND}
\label{annexe3}

La formule MOND postule une modification de la loi de Newton dans un
r\'egime de champs de gravitation plus faibles qu'une certaine
\'echelle d'acc\'el\'eration caract\'eristique $a_0$. Elle a \'et\'e
propos\'ee par Moti Milgrom~\cite{Milg1,Milg2,Milg3} il y a plus de 30
ans, et a permis de pr\'edire la myst\'erieuse corr\'elation
observ\'ee entre champ gravitationnel et mati\`ere ordinaire dans les
galaxies (loi de Milgrom).  L'id\'ee est simplement que si
l'acc\'el\'eration gravitationnelle est $g < a_0$, la loi de Newton $g
= GM/r^2$ est alors remplac\'ee par $g = \sqrt{GMa_0}/r$. Cette
modification simple produit automatiquement des courbes de rotation
plates pour les galaxies, rend compte de la loi empirique (dite de
Tully-Fisher) reliant la masse de mati\`ere ordinaire des galaxies \`a
leur vitesse de rotation, et pr\'edit une \'echelle caract\'eristique
de densit\'e de surface maximale pour les disques galactiques. La
constante d'acc\'el\'eration $a_0$ est mesur\'ee par l'ajustement de
la formule aux observations des courbes de rotation des galaxies. Elle
vaut environ $10^{-10} {\rm m}/{\rm s}^2$, et est de l'ordre de la
racine carr\'ee de la constante cosmologique en unit\'es
naturelles. La formule MOND para\^it bien \'etrange et ``os\'ee'',
mais plus de 30 ans de recherches ont confirm\'e sa valeur pour
expliquer tout un ensemble d'observations aux \'echelles
galactiques. Un nombre croissant d'astrophysiciens pensent qu'elle
constitue la clef du probl\`eme de la mati\`ere noire, au moins \`a
l'\'echelle des galaxies.

\section*{References}

\bibliography{ListeRef}

\end{document}